\documentclass[aps,preprintnumbers,superscriptaddress,twocolumn,amsmath,amssymb,floatfix,prl, nofootinbib]{revtex4-1} 
\pdfoutput=1
\usepackage{graphicx}
\usepackage[usenames,dvipsnames]{color}
\usepackage[colorlinks,bookmarks=false,citecolor=NavyBlue,linkcolor=Red,urlcolor=blue]{hyperref}
\usepackage{float}

\usepackage[export]{adjustbox}
\usepackage[normalem]{ulem}
\usepackage{lmodern}
\usepackage{graphicx}
\usepackage{comment}
\usepackage[export]{adjustbox}
\usepackage{dcolumn}
\usepackage{bm}
\usepackage{hyperref}
\hypersetup{linktocpage,colorlinks,citecolor={blue},pdfdisplaydoctitle=true,pdfpagemode=UseOutlines,bookmarksnumbered=true}

\usepackage{amssymb}
\usepackage{amsmath}
\usepackage{mathtools}

\begin{document}

\title{Creep dynamics of athermal amorphous materials: a mesoscopic approach}
\author{Chen Liu}
\affiliation{FAST, CNRS, Univ. Paris-Sud, Universit\'{e} Paris-Saclay, 91405 Orsay, France}
\author{Ezequiel E. Ferrero}
\affiliation{Centro At\'omico Bariloche, 8400 San Carlos de Bariloche, R\'io Negro, Argentina.}
\author{Kirsten Martens}
\affiliation{Universit\'e  Grenoble Alpes, CNRS, LIPHY, F-38000 Grenoble, France.}
\author{Jean-Louis Barrat}
\affiliation{Universit\'e  Grenoble Alpes, CNRS, LIPHY, F-38000 Grenoble, France.}

\begin{abstract}
Yield stress fluids display complex dynamics, in particular when driven into the transient regime between the solid and the flowing state. 
Inspired by creep experiments on dense amorphous materials, we implement mesocale elasto-plastic descriptions to analyze such  transient dynamics in athermal systems. Both our mean-field and space-dependent approaches consistently reproduce the typical experimental strain rate responses to different applied steps in stress. Moreover, they allow us to understand basic processes involved in the strain rate slowing down (creep) and the strain rate acceleration (fluidization) phases. The fluidization time increases in a power-law fashion as the applied external stress approaches a static yield stress. This stress value is related to the stress over-shoot in shear start-up experiments, and it is known to depend on sample preparation and age. By calculating correlations of the accumulated plasticity in the spatially resolved model, we reveal different modes of cooperative motion during the creep dynamics.
\end{abstract}

\maketitle

Yield-stress fluids (YSFs), such as dense emulsions, colloidal suspensions and pastes, display a rich rheological behavior that has attracted considerable attention in the last decades (see reviews by Bonn \textit{et al.}~\cite{ bonn2015yield} and Nicolas \textit{et al.}~\cite{nicolas2017deformation}).
Typically, the rheological behavior of YSFs is characterized by the flow curve measured in a stationary flowing state.
The dependency of the stationary shear stress $\Sigma$ on the applied shear rate $\dot{\gamma}$ is in many cases well described by a generalised form $\Sigma(\dot{\gamma})\approx \sigma _Y + A\dot{\gamma}^{n}$, where the prefactor $A$, the ``Herschel-Bulkley'' exponent $n$ and the yield stress $\sigma_Y$ (better referred as the \textit{dynamical} yield stress) are the relevant fitting parameters.
But it is well known that assessing the material's bulk properties at finite shear flow and in the steady state limit, does not fully account for the complex dynamics of certain YSFs.
For example, the interplay between external driving and internal aging has been shown to cause complex thixotropic behaviors \cite{bonn2009yield}, leading to non-homogeneous flow even under homogeneous driving conditions. The important challenge is then to study not only the well established flow properties in the homogeneous steady flow regime, but also the spatially resolved transient dynamics that bridges the solid response at small deformations to the flowing state at large deformations.

In recent literature, efforts have involved typically two kinds of protocol: shear start-up and creep tests \cite{chaudhuri2013onset,siebenburger2012creep,divoux2011stress,coussot2006aging}.
The first approach controls the applied strain whereas the latter records the strain rate response to a sudden stress step.
In particular, creep experiments measure the strain rate evolution in response to a fixed stress $\sigma$ applied at a given waiting time $t_w$ after sample preparation \cite{struik1978physical,siebenburger2012creep,baldewa2012delayed,leocmach2014creep,ballesta2016creep,chaudhuri2013onset,sentjabrskaja2015creep,landrum2016delayed}.
In this way, the response of the system is probed as a function of its initial age.
Such experiments reveal an intriguing behavior with two salient features:
\textbf{(i)} the strain-rate $\dot{\gamma}(t)$ in response to a stress larger than the yield stress is strongly non-linear and nonmonotonous, with a so called ``\textit{S}-shaped'' dependence  of $\dot{\gamma}(t)$ \cite{divoux2011stress,siebenburger2012creep,coussot2006aging}, including a nontrivial ``primary creep regime''  often described by a power law $\dot{\gamma}\sim t^{-\mu}$;
\textbf{(ii)} the fluidization time scale $\tau_f$ diverges when approaching the yield stress, yet in a non-universal manner.
Both  features are found to depend on sample preparation.
The experiments ultimately lead for small applied stresses to a dynamical arrest or steady creep and for sufficiently large applied stresses to steady flow or failure, depending on the material.

In this work, we reproduce and interpret the above experimental features using mesoscopic modeling approaches.
The implementations we use are  suitable for athermal amorphous systems, which constitute a large sub-class of YSFs, including foams, emulsions, physical gels and granular media \cite{bonn2015yield}; where large Peclet numbers assure that thermal fluctuations are negligible compared with mechanical fluctuations induced by the response to an external driving. 
Note that the material mechanical properties and dynamics will always depend on its preparation protocol and subsequent waiting period prior to deformation; during which slow temperature-dependent processes such as glassy relaxation are indeed relevant.
In any case, our approach complements previous studies addressing creep in systems for which thermal fluctuations are also important during the dynamics, based on the soft glassy rheology model \cite{petersgr1998, fielding2000aging, moorcroft2013criteria, merabia2016creep} or mode-coupling theory \cite{mctreho2002, brader2009glass}.
We work in particular with mean-field~\cite{LiuPRL18} and spatially-resolved versions of elasto-plastic models;
where the flow of YSFs results from the interaction among local plastic events triggered by external driving.
It has been shown that such models account for various properties of YSFs \cite{Argon79, PicardPRE05, NicolasPRL13, PuosiPRE14, LiuPRL16, albaret2016mapping, boioli2017shear}.
While previous formulations consider a fixed strain rate as a control parameter \cite{NicolasEPL14, LiuPRL16, PicardEPJE04, HL, agoritsasEPJE15}, we extend those models here to describe the constant imposed stress condition relevant for creep.
With that, we reproduce the experimental features \textbf{(i)} and \textbf{(ii)} explained above.
Interestingly, we find the same fluidization time dependence on initial aging as \citet{siebenburger2012creep}.
Furthermore,  we show that the divergence of the fluidization time is described by a power-law relation when the distance to yield is measured with respect to an age-dependent static yield stress $\sigma^s_Y$ larger than the dynamical yield stress $\sigma_Y$ itself.
As in experiments, the resulting power-law exponent appears to be non-universal.  

\section*{Methods}

\subsection*{Mesoscopic elastoplastic model}

Our spatially-resolved approach is extended from a previous version used to describe  steady state flows \cite{PicardEPJE04, MartensPRL11, NicolasEPL14, LiuPRL16}.
It consists in a square lattice  where a node $(ij)$  represents a typical cluster of particles undergoing plastic events \cite{Argon79} and $i$, $j$ the discretized coordinates along the $x$ and $y$ directions respectively. Once a reference state is set, each node is associated with a local plastic deformation $\gamma^{pl}_{ij}$ which is, in general, heterogeneous.
In addition, a node can develop an elastic strain $\gamma^{el}_{ij}$ associated with a local stress $\sigma_{ij}=\mu\gamma^{el}_{ij}$.
The local stress is composed of two parts, $\sigma_{ij}=\sigma^{EXT}+\sigma^{INT}_{ij}$, where $\sigma^{EXT}$ is the externally applied uniform stress, and $\sigma^{INT}_{ij}$ encodes the stress fluctuations resulting from the interactions between plastic heterogeneities, more precisely
\begin{eqnarray}\label{simga_int}
 \sigma^{INT}_{ij}=\mu\sum_{i'j'}G_{ij,i'j'}\gamma^{pl}_{i'j'}\;.
\label{eq:sigma_int}
\end{eqnarray}
The interaction kernel $G$ is of the Eshelby's type \cite{Eshelby57} as described in \cite{PicardEPJE04}, plus an homogeneous part $1/N$ with $N$ the system size, so that the integral over space of the internal stress caused by any plastic strain field is null.
Thus $\sigma^{INT}_{ij}$ describes the local stress fluctuations  in a macroscopically stress free state. Applying a macroscopic stress amounts to  shifting uniformly the local stress without altering internal fluctuations.

Besides, each node alternates between a local plastic state and a local elastic state by switching a local state variable $n_{ij}$ between $1$ and $0$,respectively.
Explicitly when $|\sigma_{ij}|$ is larger than $\sigma_c$, the site becomes plastic ($n_{ij}=0\longrightarrow n_{ij}=1$) at a rate $1/\tau_{pl}$ and becomes elastic again ($n_{ij}=1\longrightarrow n_{ij}=0$) at a rate $1/\tau_{res}$ \cite{PicardEPJE04,PicardPRE05}, i.e.
\begin{equation}
n_{ij}(t): \;\; 0\xrightleftharpoons[\forall \sigma,\; \tau^{-1}_{res}]{|\sigma|>\sigma_c,\;\tau^{-1}_{pl}} 1
\end{equation}
The local dynamics is expressed as
\begin{eqnarray}\label{local_dym}
 \frac{d}{dt}\gamma^{pl}_{ij}=n_{ij}\frac{\sigma_{ij}}{\mu\tilde{\tau}}=n_{ij}\frac{\sigma^{INT}_{ij}+\sigma^{EXT}}{\mu\tilde{\tau}}\;.
\end{eqnarray}

We set in our simulation $\tilde{\tau}=\tau_{res}=\tau_{pl}=1$, $\sigma_c=1$ and $\mu=1$.
The model described above is a reformulation of the model in \cite{PicardEPJE04} but allowing us to model a stress controlled protocol.
To summarize, Eq.~(\ref{simga_int}) and Eq.~(\ref{local_dym}) forms a closed stochastic dynamical system governing the evolution of  the plastic strain field $\gamma^{pl}_{ij}(t)$.
Given the initial condition defined by $\gamma^{pl}_{ij}(t=0)$ and the imposed stress $\sigma^{EXT}$, we simulate the system and measure the macroscopic plastic strain $\langle\gamma\rangle^{pl}(t)\hat{=}\sum_{ij}\gamma^{pl}_{ij}(t)/N$, from which the strain rate response $\langle\dot{\gamma}\rangle^{pl}(t)$ results directly.\par

\subsection*{Mean-field approach for creep dynamics}

The probability distribution function $P(\sigma,t)d\sigma$ gives the fraction of nodes with a local stress in  $[\sigma,\sigma+d\sigma]$ at time $t$.  Our mean-field approach to approximating the time evolution of this probability distribution for a typical site  is inspired by the H\'ebraud-Lequeux model \cite{HL} and thus belongs to the class of athermal local yield stress models \cite{agoritsasEPJE15, EliArXiv16}. One should note that for the derivation of this model, we assume several strong simplifications with respect to the spatial elasto-plastic description, such that all comparisons can only be done on a qualitative level. As we will discuss later the detailed aspects of the creep curve depend  on these simplifications whereas other more general features, like power-law scalings, appear to be very general. 
Our mean-field approach describes the dynamics of the distribution $P(\sigma,t)$, and differs from the original model \cite{HL} by further taking into account a strain rate that may vary in time  $\dot\gamma(t)$:
\begin{eqnarray}
\partial_t P(\sigma,t) &=& - \frac{1}{\tau}\theta(|\sigma|-\sigma_c)P(\sigma,t) + \Gamma(t) \delta(\sigma) \nonumber\\
 && - G_0 \dot{\gamma}(t)  \partial_\sigma P(\sigma,t)  + D(t) \partial_\sigma^2 P(\sigma,t)\;.
\label{eq:hl}
\end{eqnarray}
The first term on the right hand side describes local yielding with rate $1/\tau$ if the local stress exceeds a yield stress $\sigma_c$ ($\theta$ denotes the usual Heaviside distribution) and the second term is the corresponding gain term accounting for a instantaneous complete relaxation of the stress. Here $\delta(\sigma)$ is the Dirac distribution, and $\Gamma(t)$ is the rate of plasticity
\begin{equation}
\Gamma(t)= \frac{1}{\tau} \int_{|\sigma|>\sigma_c} d\sigma P(\sigma,t)\ .
\end{equation}
The third term  accounts for the local elastic response with shear modulus $G_0$. The last term in Eq.~(\ref{eq:hl}) is a mean-field approximation for interactions between different macroscopic regions presented in the spatially resolved model. This term describes in an effective way the stress fluctuations caused by the elastic response to surrounding yielding events as a diffusive stochastic process with a time-dependent diffusion constant $D(t)$ proportional to the rate of plasticity $\Gamma(t)$:
\begin{equation}
 D(t) = \alpha \Gamma(t)\ .
\end{equation}
This approximation by a diffusive process assumes that the "kicks" received by a typical site are uncorrelated and of finite variance, an approximation to the true  dynamics of elastoplastic models that  has been improved, in the quasistatic limit of vanishing strain rate, by the work of Lin and Wyart \cite{LinWyartPRX2016} taking into account the fact that these kicks have a broad distribution. We are not aware of any similar extension to finite strain rates, and in this work will remain at the simpler level of description. 

Driving the system with a constant stress, $\int d\sigma \sigma P(\sigma,t) = constant$, leads

by integration of \ref{eq:hl} to a self consistent determination of the strain rate as  
\begin{equation}
\dot{\gamma}(t) = \frac{1}{\tau G_0}\int_{|\sigma|>\sigma_c} d\sigma \sigma  P(\sigma,t).
\label{eq:selfconsistent}
\;\end{equation}
Note that this strain rate corresponds physically to the rate of total plastic strain released by the yielding sites, each site releasing $\sigma/G_0$, 
Equations \ref{eq:hl} and \ref{eq:selfconsistent} can be solved numerically starting from a given initial condition $P_0= P(\sigma,t=0)$.

\subsection*{Initial condition and aging}

In the mean-field approach, the initial condition is fully defined by $P_0(\sigma)=P(\sigma,t=0)$.
To represent the quenched state of a system before applying the step stress, we consider a distribution of zero mean $P^I_0(\sigma)$ that describes the internal stress fluctuations of a system in a macroscopically stress free state.
This distribution will be instantaneously shifted to the desired value of the imposed stress $\sigma^{EXT}$ at the beginning of the creep protocol to mimic the application of a stress step, i.e. $P_0=P^I_0(\sigma - \sigma^{EXT})$.
In the spatially resolved elasto-plastic model, the initial state is defined by $\gamma^{pl}_{ij}(t=0)$. 
Using  Eq.~(\ref{eq:sigma_int}), $\gamma^{pl}_{ij}(t=0)$ can be converted to a field of internal stress fluctuations $\sigma^{INT}_{ij}(t=0)$ from which a zero mean distribution, such as $P^I_0(\sigma^{INT})$, can be constructed.
Then the distribution of local stresses at the onset of creep experiments can also be described as an instantaneous shift of the zero mean distribution with the imposed step stress $\sigma^{EXT}$, such as $P^I_0(\sigma - \sigma^{EXT})$.
In practice, we first choose a specific form of $P^I_0(\sigma)$, then convert it back to a random realization of $\gamma^{pl}_{ij}(t=0)$ and the creep test is simulated by evolving the model under a fixed value of $\sigma^{EXT}$.
Once an initial condition is prepared, we numerically integrate the dynamic equations using an explicit Euler method.

In principle we should consider only distributions $P^I_0(\sigma)$ with a compact support in both models, i.e. $P^I_0(\sigma>\sigma_c)=0$, so that the system does not evolve until the external load is applied.
Hence, our models do not explicitly resolve the aging dynamics, but we mimic the role of aging by using different choices of the initial condition.
In a first approach, we assume for the distribution $P^I_0(\sigma^\mathrm{int})$ a Gaussian shape\footnote{Strictly speaking this initial condition may violate the compact support, but in practice the standard deviations studied are small enough such that the statistical weight beyond $\sigma_c$ can be regarded as negligible.} centered at zero \cite{srolovitz1981radial}.
The only parameter is the standard deviation $s_d$, characterizing the level of residual heterogeneity in the amorphous system.
As more relaxed systems display a less prominent ``Boson peak'', indicative of a better homogeneity of the elastic properties \cite{duval2006physical}, 
we assume that relaxation is also reducing the width of the stress distribution.
Thus a $P^I_0(\sigma^\mathrm{int})$ with a smaller $s_d$ corresponds to a more relaxed system, and we take $1/s_d$ as an indirect measure of the age.
Interestingly, the standard deviation of our distribution can be formally linked to the aging parameter in the lambda-model for thixotropic materials \cite{wei2016quantitative,armstrong2016dynamic}, as discussed in the Supplementary Material of Ref.~\cite{LiuPRL18}.

\section*{Results}

\subsection*{Flow curves}

\begin{figure}[t]
\begin{center}
\includegraphics[width=\columnwidth, clip]{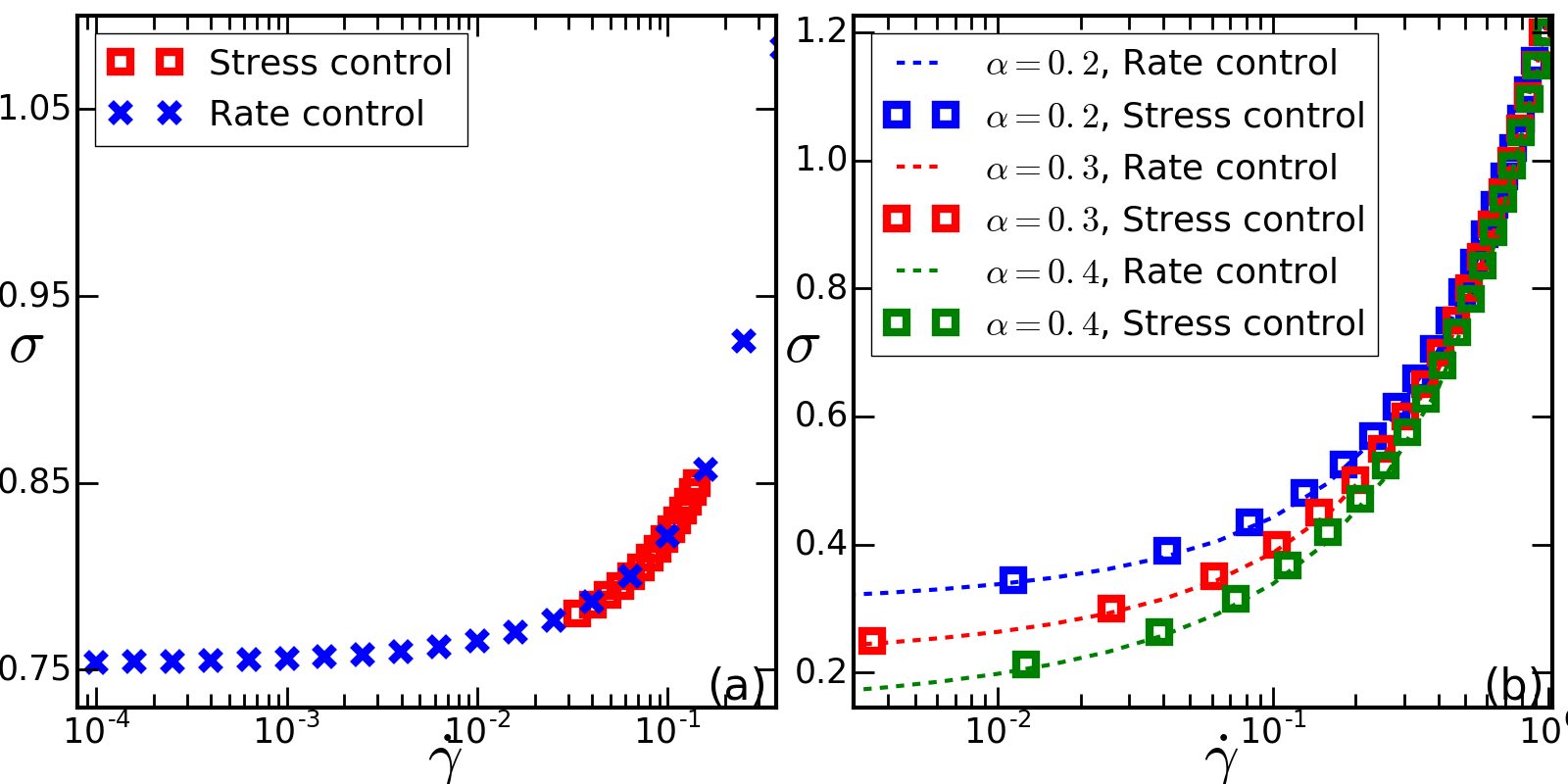}
\end{center}
\caption{
Check of the  consistency between different protocols.
Dashed line: flow curves obtained by strain rate control protocol.
Squares: flow curves obtained by stress control protocol.
(a) Spatially-resolved elasto-plastic model.
(b) Mean-field model.
}
\label{fig:consistent}
\end{figure}
Let us start, prior to the study of the creep behavior, by comparing the flow curves obtained from the stress control protocol here presented to those obtained from the strain rate control protocol explained elsewhere \cite{PicardEPJE04,HL}.
In the stress control protocol, we set the stress by choosing an arbitrary initial condition among those with the desired stress value.
The system evolves with a stress-preserved dynamics and reaches a steady flow regime at large times when the memory of the initial condition is completely erased. 
We then average $\dot{\gamma}(t)$ over a large time window in the steady state.
For both the spatial model and the mean-field approach, the comparison shown in Fig.~\ref{fig:consistent} reveals a good consistency between the two types of protocols.
The dynamic yield stress $\sigma_Y$ of the spatially-resolved model is estimated to be $\sim 0.7536$.
The dynamic yield stress $\sigma_Y$ of the mean-field model is a decreasing function of the mechanical coupling strength $\alpha$, as explained in Ref.~\cite{agoritsasEPJE15}. 

\subsection*{Creep curves}

In the following, we use  exclusively the stress-controlled protocol. 
An initial condition with an imposed  stress is chosen, corresponding to the application of a step stress on a relaxed sample in experiments.

\begin{figure}[th]
\begin{center}
\includegraphics[width=\columnwidth, clip]{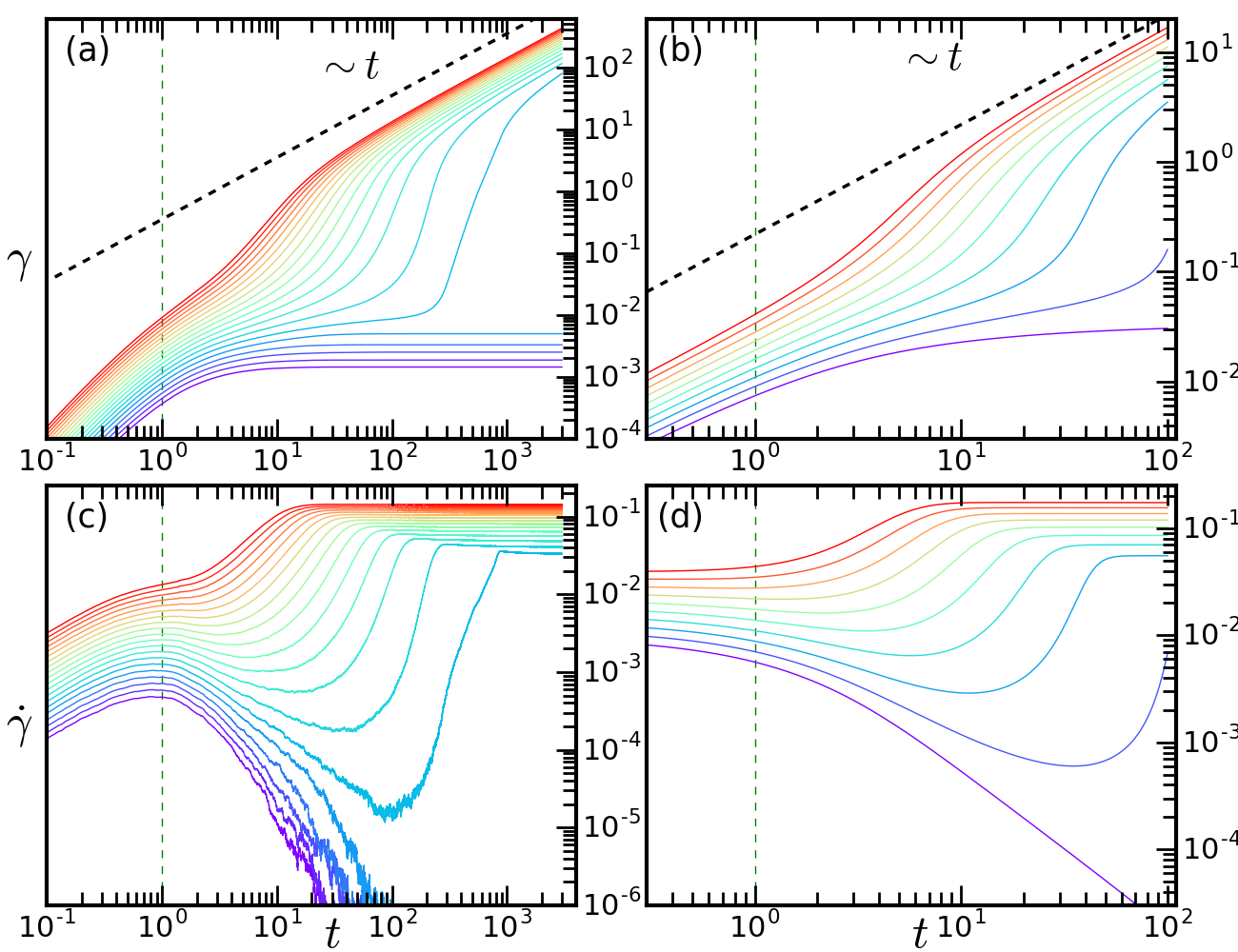}
\end{center}
\caption{
{\it Creep behavior at different imposed stresses for the same initial relaxation (aging).}
Left and right columns show respectively data produced by the elasto-plastic and mean-field models.
The upper row shows the strain time series and the bottom row shows the corresponding strain rate time series.
Elasto-plastic model: $s_d\approx 0.083$, values of $\Delta\sigma$ from purple to red (bottom to top) $0.005, 0.01, 0.015, \ldots, 0.1$.
Mean-field model: $\alpha=0.4$, $s_d=0.32$, values of $\Delta\sigma$ from purple to red (bottom to top) $0.1, 0.12, 0.14, \ldots, 0.28$. }
\label{fig:creep1}
\end{figure}

\begin{figure}[th]
\begin{center}
\includegraphics[width=\columnwidth, clip]{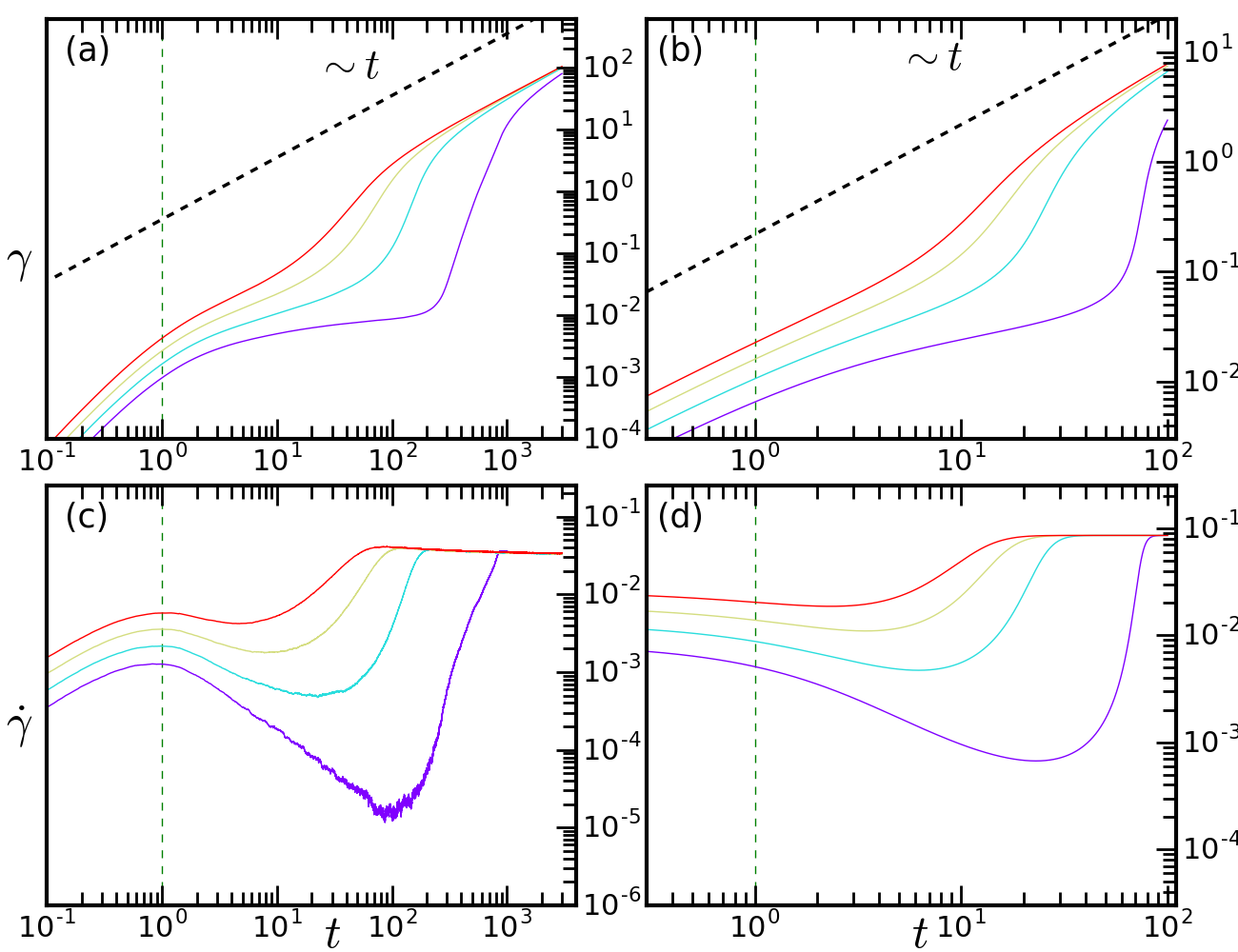}
\end{center}
\caption{
{\it Creep behavior for different initial relaxation at the same imposed stress.}
Left and right columns show  data produced, respectively, by the elasto-plastic and mean-field models.
The upper row shows the strain time series and the bottom row shows the corresponding strain rate. 
Elasto-plastic model: $\Delta\sigma=0.03$, $s_d$ values from purple to red (bottom to top) $0.083, 0.089, 0.096, 0.104$. Mean-field model: $\Delta\sigma=0.18$, $s_d$ from purple to red (bottom to top): 0.28,0.3,0.32,0.34
}
\label{fig:aging}
\end{figure}

Typical responses of $\dot{\gamma}(t)$ and $\gamma(t)$ for the two models just after the application of a step stress are shown in Fig.~\ref{fig:creep1} and Fig.~\ref{fig:aging}.
All these curves are obtained with an imposed stress larger than the dynamical yield stress, i.e. $\Delta \sigma\equiv \sigma ^{EXT}-\sigma_Y>0$. 
The fact that the two models differ in their behaviors of $\dot{\gamma}(t)$ for $t\lesssim 1$ is due to the different ways in which they describe the plasticity, namely a viscous relaxation in the spatially resolved model compared to an instantaneous one in the mean field approach.
Beyond a microscopic time scale $t_{mic} \simeq 1$, up to which they depend on the details of the different dynamics, we notice that the strain responses of the two models behave qualitatively in the same way. Already at early times after $t_{mic} \simeq 1$, the strain rate $\dot{\gamma}$ monotonically decreases for the smallest $\Delta\sigma > 0$ cases and correspondingly the strain $\gamma$ reaches a plateau. 
This implies that even when the applied stress step is above the dynamic yield stress, a quenched system submitted to a creep test may not yield to a flowing state. This is one of the main differences with fixed shear rate protocols, where the system is always forced to yield. We will come back to this point later.
For a larger imposed stress, instead, both the spatial and mean-field models reproduce the characteristic \textit{S}-shaped curve for $\dot{\gamma}(t)$ observed in experiments \cite{coussot2006aging,siebenburger2012creep,divoux2011stress}.

For a fixed initial aging level ($s_d$ fixed in Fig.~\ref{fig:creep1}), before entering the  plateau of the steady-state regime, $\dot{\gamma}(t)$ displays a creep regime where the strain rate decreases with time in an apparent power-law fashion until it reaches a minimum.
Within this creep regime, the accumulated strain $\gamma(t)$ shows a sub-linear increase in time.
After the minimum in $\dot{\gamma}$, the system enters a fluidization regime where the strain rate speeds up toward the steady-state, and correspondingly the accumulated strain $\gamma(t)$ increases super-linearly to reach the linear regime of a steady flowing state. As we further increase the stress, the extent of the creep regime decreases, until it eventually disappears and the system enters directly the fluidization regime after $t_{mic}$.

A similar effect is found for a given applied stress when we vary the initial aging level (Fig.~\ref{fig:aging}).
The duration of the creep regime decreases when increasing $s_d$ (decreasing age), up to the point where it disappears for large enough $s_d$, and all curves reach the same steady strain rate.
This indicates that, at a given applied stress, a less relaxed system is more likely to be fluidized. These dependences on the applied stress and the initial relaxation are reminiscent of the creep tests performed in bentonite suspensions \cite{coussot2006aging} and colloidal hard sphere systems \cite{siebenburger2012creep}.

Several works suggest a power law slowing down $\dot{\gamma}\sim t^{-\mu}$ in the creep regime \cite{divoux2011stress,ballesta2016creep,landrum2016delayed,Roy2018,miguel2002}. Our data can be indeed fitted with such a power law for a modest range of the parameters $(s_d, \sigma^{EXT})$.
By doing so in curves where we can fit at least one decade of power-law, we observe that the exponent $\mu$ decreases with increasing applied stress and decreasing initial aging (increasing $s_d$).
The values of $\mu$ extracted from the two models are similar and vary from $0.6$  to $1.2$ (fits not shown here) depending on $s_d$ and $\sigma ^{EXT}$, a range comparable to those reported in experiments \cite{divoux2011stress,ballesta2016creep,landrum2016delayed}.

Note that, in order to produce a comparable time dependence of the strain rate, significantly different values of the control parameters $\Delta \sigma$ and $s_d$ are used as input in the two different models, as shown in Fig.~\ref{fig:creep1} and Fig.~\ref{fig:aging}. 
In principle some of the differences in the creep response could result from finite size effects in the spatial model and discretization effects in the mean-field description. But here, we took care that for the parameter range studied, these effects are not relevant. Instead, to understand this discrepancy, one has to recall that the rules for the local plastic deformation in the two models are very different. In the mean-field model the local stress release is instantaneous and complete whereas the spatial model implements a duration of events and only a partial local stress relaxation. This is also the origin of different behaviors of the two models in their $\dot{\gamma}(t)$ for $t\lesssim t_{mic} \simeq 1$. Before the vertical dashed line (Fig.\ref{fig:creep1} and Fig.\ref{fig:aging}), the strain rate $\dot{\gamma}(t)$ from the elasto-plastic model increases linearly (thus $\gamma(t)$ increases quadratically) until $t\approx t_{mic}$, while the strain rate from the mean-field model begin with a finite value and varying significantly only after $t_{mic}\simeq 1$. Another obvious difference is the interaction kernel of the spatial model, leading to a spatially correlated dynamics. As shown before, the flow curves do not match and in the mean-field model, they depend on the value of $\alpha$, and consequently so do the creep curves. For these reasons we can only hope to qualitatively reproduce the creep curves with the mean-field model. Interestingly,  however there are other quantities like the fluidization time dependence
on the imposed stress that are still quantitatively comparable and thus point towards more general behavior. We will discuss this point in detail in the following section.

\subsection*{Fluidization time}

We now discuss the relation between the fluidization time scale and the distance of the applied stress to the dynamical yield stress $\Delta\sigma =\sigma^{EXT}-\sigma _Y$.
Two characteristic times can be recognized in the evolution of $\dot{\gamma}$, at least for intermediate values of the applied stress.
We call $\tau_m$ the time where the minimum of $\dot{\gamma}$(t) occurs, after the creep regime, and define $\tau_f$ as the inflection point of $\dot{\gamma}(t)$ in the fluidization regime before entering the steady state.
Following \cite{divoux2011stress}, we choose $\tau_f$ to characterize a fluidization time scale.
Note that $\tau_f$ is always defined, even in the absence of a creep regime (e.g., for large values of $\Delta\sigma$) where we can still recognize a fluidization phase with an acceleration of $\dot{\gamma}$, and it will characterize the typical time needed to reach the steady flow.

\begin{figure}[th]
\begin{center}
\includegraphics[width=\columnwidth, clip]{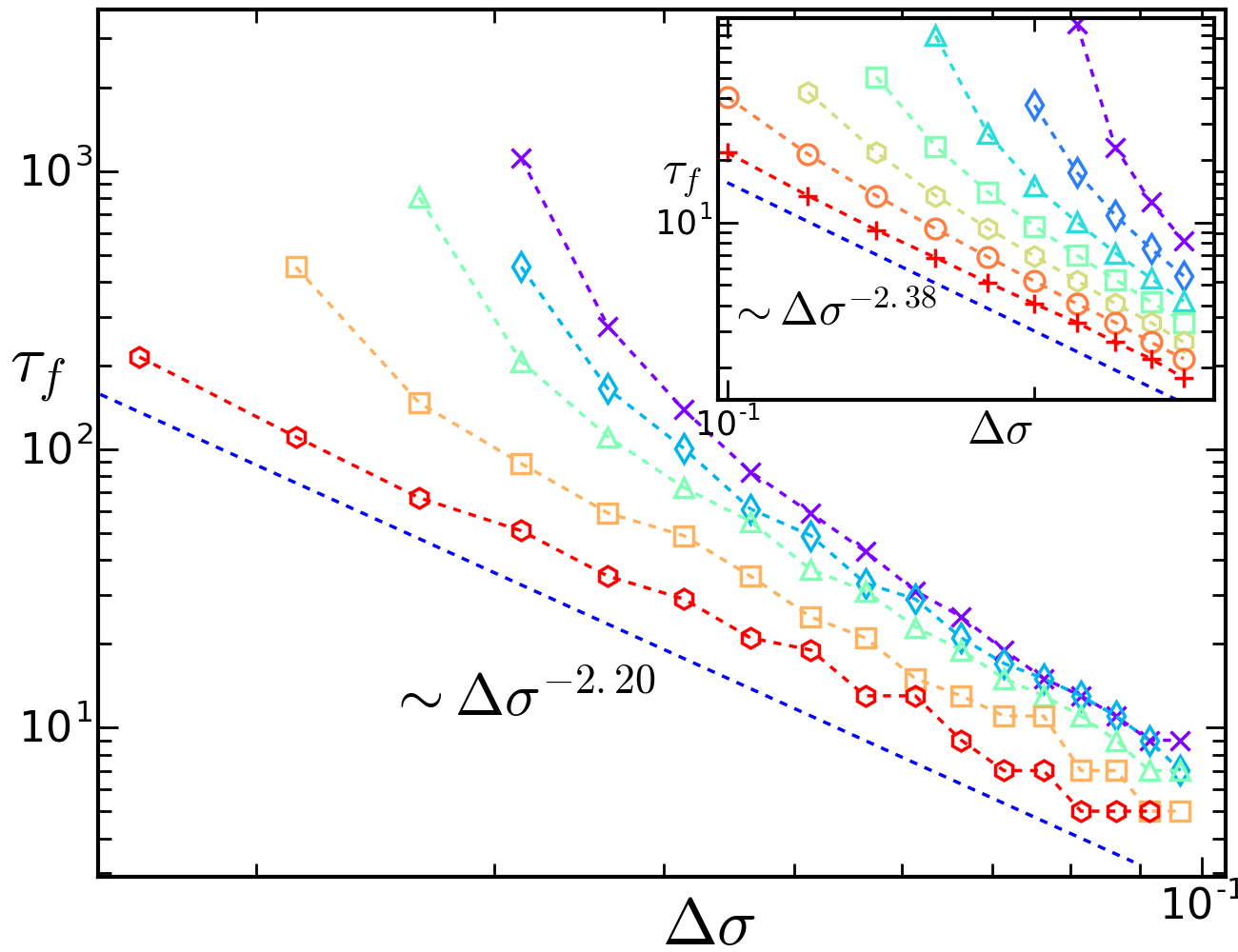}
\end{center}
\caption{
{\it Fluidization time $\tau_f$ vs $\Delta\sigma$ for different initial relaxation levels.}
Main figure: Elasto-plastic model, $s_d$ increases from the top to the bottom ($s_d = 0.078, \; 0.081,\; 0.083,\; 0.089,\; 0.096$).
Inset: Mean-field model with $\alpha=0.3$, $s_d$ increases from the top to the bottom ($s_d=0.22,\;0.24,\; 0.26,\;...,\;0.34$).
}
\label{fig:tf}
\end{figure}

Fig.~\ref{fig:tf} shows the $\tau_f$ dependence on $\Delta\sigma$ for different initial aging levels.
Experimental results on a carbopol microgel \cite{divoux2011stress} suggest a power law dependence $\tau_f\sim \Delta\sigma ^{-\beta}$ with $\beta$ measured from $2$ to $8$ depending on sample preparation.
This behavior must be distinguished from studies on thermal systems \cite{merabia2016creep,lindstrom2012structures,sprakel2011stress,gibaud2009shear} which suggest an exponential relation instead.
Our results show, especially when $s_d$ is small (well relaxed systems), a convexity of the curves indicating that the fluidization time increases  faster than a power law as $\Delta\sigma$ approaches zero.
As $s_d$ becomes larger, $\tau _f(\Delta\sigma)$ becomes more power-law-like.
Fig.~\ref{fig:tf} also shows that  more relaxed systems display a  stronger  increase in $\tau_f$ for decreasing $\Delta\sigma$.
This aging dependence of $\tau _f(\Delta\sigma)$ qualitatively agrees with the experimental results of \citet{divoux2011stress}. 

\begin{figure}[bh]
\begin{center}
\includegraphics[width=\columnwidth, clip]{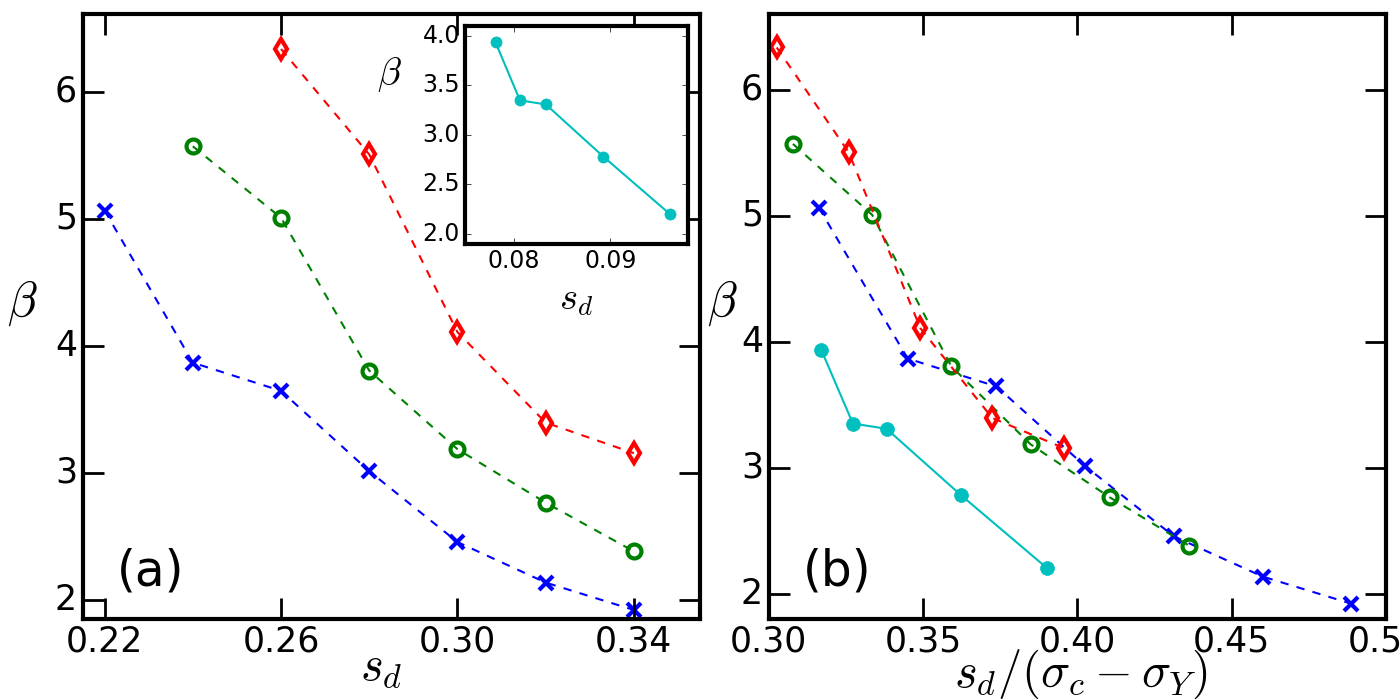}
\end{center}
\caption{
(a) Exponent $\beta$ vs standard deviation $s_d$. Main figure: mean-field model. Blue: $\alpha=0.2$, Green: $\alpha=0.3$, Red: $\alpha=0.4$. Inset: Spatially resolved model. (b) Exponent $\beta$ vs $s_d/(\sigma_c-\sigma_Y)$. The labels for different curves are the same as in (a).
}
\label{fig:beta}
\end{figure}

Despite the different functional forms when changing $s_d$, to quantify the dependence of $\tau _f(\Delta\sigma)$ on aging, we define for each curve a value of $\beta$.
When appropriate (large $s_d$, as discussed above) we directly fit a power law $\tau _f = A \Delta\sigma^{-\beta}$, finding in both the spatial and mean-field models consistent values ($\beta\approx 2.2$ and $2.38$, respectively), lying in the range of experimental results.
For initially  well relaxed systems (small $s_d$), we define an effective $\beta = \max \frac{d\ln \tau_f}{d\Delta \sigma}$, that equally carries information on how fast the fluidization process slows down with decreasing stress.
Results are shown in Fig.~\ref{fig:beta}(a).
We see a systematic decrease of $\beta$ with $s_d$ for both models.
Moreover, the estimated values of $\beta$ cover the same range (from $2$ to $8$) found in carbopol microgels \cite{divoux2011stress,grenard2014timescales}.
We also notice that, for the mean field model the exponent $\beta$ depends on the mechanical coupling strength $\alpha$.
A larger mechanical coupling strength yields a larger value of $\beta$ for the same $s_d$. It has been proposed in earlier works \cite{bocquet2009kinetic} that the value of $\alpha$ should depend on the specific form of the interaction kernel, and within simulations of particle based models it has been found that $\alpha$ typically lies in the range\cite{PuosiSoftMatter2015} of 0.26 to 0.33, which is consistent with the range of $\alpha$ of the mean-field model studied here.

\subsection*{Rationalization of the creep dynamics}

In this section, we attempt to rationalize the behavior of the global observables reported  above by analyzing in detail the evolution of the stress probability distribution.
Although the two models seem quite different in their formulation, the underlying physical process are quite similar.
For example, local plasticity in the mean-field model consists of a total release of local stress and a sudden return to the elastic state, which can be viewed as the local plasticity in the elasto-plastic model in the limit $\tau_{res}\rightarrow 0$ and $\tilde{\tau}\rightarrow 0$.
In the elasto-plastic model, the stress released by plastic events is re-distributed according to the interaction kernel through Eq.~(\ref{eq:sigma_int}) keeping the average stress constant but broadening the distribution $P(\sigma,t)$ with an amplitude roughly proportional to the rate of plastic events.
This effect of stress redistribution is reasonably approximated in the mean-field model by the diffusive term in Eq.~(\ref{eq:hl}) with the diffusion coefficient proportional to the plastic activation rate.
Although the more realistic elasto-plastic model contains more information, the simpler mean-field model already captures the key ingredients needed to understand the underlying mechanism of creep in athermal amorphous systems.

We therefore focus on the mean-field model and, to gain insights into the complex dynamics during creep tests, proceed to analyse the complex nonlinear behaviour of  Eq.~(\ref{eq:hl}) in a qualitative manner.
The time evolution of  $P(\sigma,t)$ is driven by the existence of a population of sites in an unstable state, ($P(\sigma>\sigma_c)$). 
Actually, by integrating Eq.~(\ref{eq:hl}) beyond $\sigma_c$, one obtains
\begin{equation}\label{eq:evolv_Gamma}
\tau \dot{\Gamma}(t)\approx -\Gamma(t) + \Gamma(t)\bigg[\sigma_c|P(\sigma_c)| + \alpha |\partial_{\sigma}P(\sigma_c)|  \bigg]
\end{equation}
\noindent using that the negative part beyond $-\sigma_c$ is negligible for positively imposed stress and $\dot{\gamma}(t) \approx \frac{\sigma_c}{G_0} \Gamma(t)$, (since $P(\sigma)$ weights little and decreases fast beyond $\sigma_c$).
When the population of unstable sites $\tau\Gamma$ is non-zero, it decreases  exponentially due to the loss term of plastic activation represented by the first term on the r.h.s of Eq.~(\ref{eq:evolv_Gamma}). 
At the same time, it is supplied by two comparable fluxes represented by the last term in Eq.~(\ref{eq:evolv_Gamma}): the stress drift due to the elasticity and the stress diffusion due to stress redistribution.
These fluxes are both proportional to the unstable population.
In particular, the drift and diffusion induced fluxes are respectively proportional to $P\big|_{\sigma = \sigma_c}$ and $\partial_{\sigma}P\big|_{\sigma=\sigma_c}$. 

Let us first consider two extreme cases with $\sigma^{EXT}>\sigma_Y$, where a steady flowing state exists according to the flow curve.
If the standard deviation of the initial Gaussian $P_0(\sigma)$ is large enough, the supply of the unstable sites is as important as the plastic activation.
Thus, the supply and the loss rapidly reach a situation where they compensate each other.
This corresponds to the curves with no creep regime at high imposed stresses in Figs.~\ref{fig:creep1} and \ref{fig:aging}.
On the other hand, if the standard deviation of the initial distribution $P_0(\sigma)$ is very small, not only a small portion of the population is unstable but also the values of $P\big|_{\sigma = \sigma_c}$ and $\partial_{\sigma}P\big|_{\sigma=\sigma_c}$ are close to zero. The term of supply is then negligible compared to the  loss term in Eq.~(\ref{eq:evolv_Gamma}).
As a consequence, drift and diffusion are as weak as the unstable population at the beginning and become even weaker as the unstable population decreases exponentially.
The strain rate rapidly decreases, the system gets eventually stuck in a configuration where all sites are below $\sigma_c$ and the flow stops.
This situation corresponds to the curves with vanishing strain rate in Fig.~\ref{fig:creep1}.
Note that this situation can be observed in experiments and simulations with fixed stress protocols \cite{bonn2015yield} even if the applied external stress is larger than the dynamic yield stress $\sigma_Y(\alpha)$.

The above analysis raises the question of the evolution of $P(\sigma,t)$ that causes the transition from the creep regime to the fluidization regime and eventually the steady flow, observed for intermediate values of $s_d$.
It further suggests that $\tau_f$ can diverge \textit{before} $\Delta\sigma$ tends to zero (see Fig.~\ref{fig:tf}).
For a  well relaxed system (small $s_d$), we can therefore introduce a static yield stress $\sigma^s_Y$ defined as the minimum stress needed to fluidize a system at rest.
$\sigma^s_Y$ will depend on the initial state and will be larger than the dynamical yield stress $\sigma_Y$.
In an extreme situation where $s_d=0$, one should apply $\sigma^\mathrm{EXT}\geq\sigma^s_Y=\sigma_c$ to make the system flow.
This is consistent with previous studies on transient dynamics that report  an age dependent overshoot in the stress-strain curve \cite{fielding2000aging,rottlerrobbins,rodney2011potential,varnik2004study}. 
Actually, the stress overshoot in the zero strain rate limit is closely related with $\sigma^s_Y$.
A comparison between the two is discussed later on. 

\begin{figure}[th]
\begin{center}
\includegraphics[width=\columnwidth, clip]{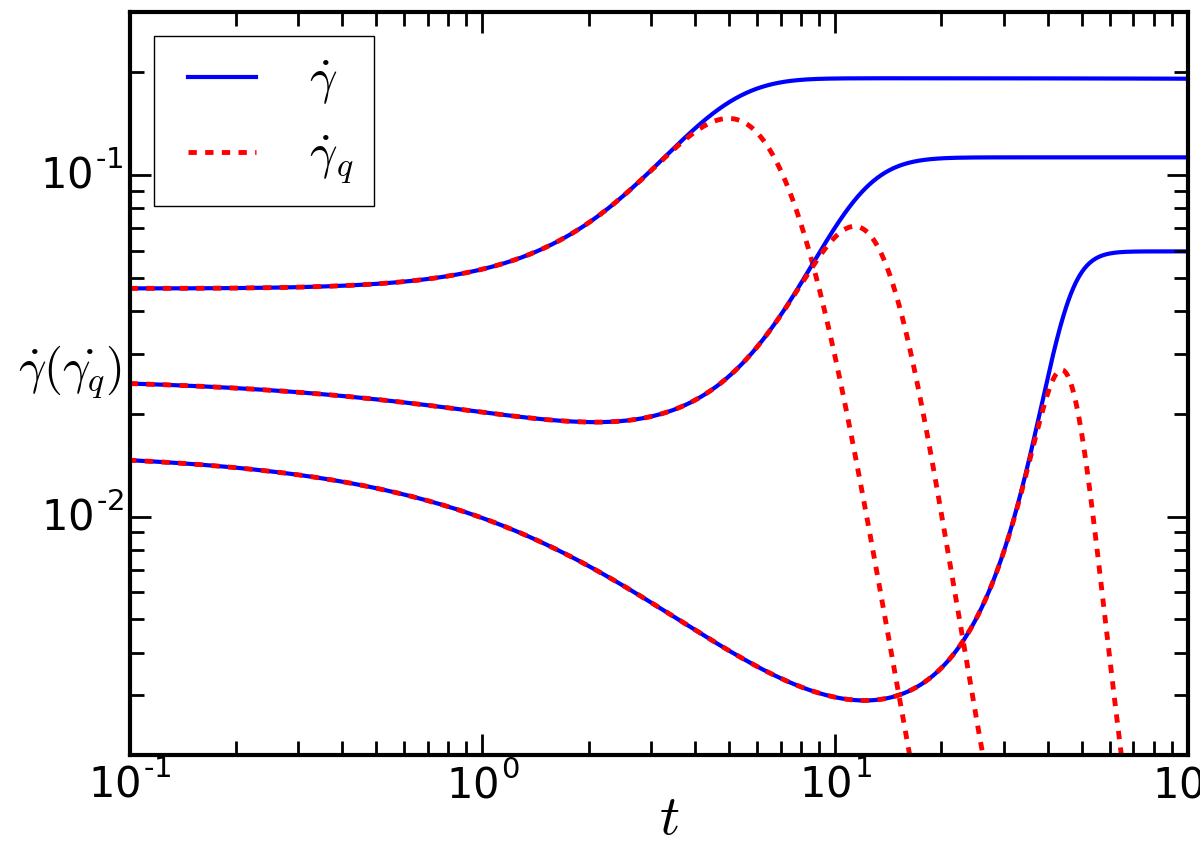}
\end{center}
\caption{
Comparison between $\dot{\gamma}_q(t)$ (red dashed curves) generated by Eq.~(\ref{eq:hl_pq}) and $\dot{\gamma}(t)$ (blue solid curves) generated by Eq.~(\ref{eq:hl}) for three different applied stresses by which we have different extents of creep regimes. 
}
\label{fig:shearrate_q}
\end{figure}

To gain a better understanding of the initial evolution of the strain rate, we study the full dynamics in the early regime -such that mesoscopic blocks have been activated at most once- by setting $P = P_q + P_a$, with $P_q$ referring to the sites that have never been activated and $P_a$ to those activated once.
Thus $P_q(\sigma,t=0)=P_0(\sigma)$, $P_a(\sigma,t=0)=0$ and the distributions obey:
\begin{eqnarray}\label{hl_pq}
\partial_t P_q(\sigma,t) &=& - G_0 \dot{\gamma}_q(t)  \partial_\sigma P_q(\sigma,t) + D_q(t) \partial_\sigma^2 P_q(\sigma,t) \nonumber\\
 && - \frac{1}{\tau}\theta(|\sigma|-\sigma_c)P_q(\sigma,t) \;.
\label{eq:hl_pq}
\end{eqnarray}
\begin{eqnarray}\label{hl_pa}
\partial_t P_a(\sigma,t) &=& - G_0 \dot{\gamma}_q(t)  \partial_\sigma P_a(\sigma,t) + D_q(t) \partial_\sigma^2 P_a(\sigma,t) \nonumber\\
 && +\Gamma_q(t)\delta(\sigma) \;.
\label{eq:hl_pa}
\end{eqnarray}
where $\dot{\gamma}_q(t)$, $\Gamma_q(t)$ and $D_q$ are defined as above, with $P$ replaced by $P_q$.
We note that Eq.~(\ref{eq:hl_pq}) and Eq.~(\ref{eq:hl_pa}) approximate the full dynamics, ignoring the possibility of multiple activation.
As a result they will always lead to a vanishing strain rate at long times, $\dot{\gamma}_q(t)\rightarrow 0$.
However, the comparison between this approximation and the full solution will give us insights into the time range over which the initial condition influences the fluidization process.

In Fig.~\ref{fig:shearrate_q}, we compare the approximate solution $\dot{\gamma}_q(t)$ obtained by solving Eqs.\ref{eq:hl_pq} and \ref{eq:hl_pa} with the full solution $\dot{\gamma}(t)$, for the same initial settings $(s_d,\sigma^{EXT})$. We used three different values of the applied stress to have different extents of the creep regime. Fig.~\ref{fig:shearrate_q} confirms that in all situations, up to the mid-fluidization regime, $\dot{\gamma}_q(t)$ and $\dot{\gamma}(t)$ are in good agreement, indicating that the dynamics governing the creep regime is dominated by the sites undergoing their very first activations. The curves also show that multiple plastic activations must come into play for the crossover from the  fluidization regime to the steady flow to take place. We can conclude that the existence of a creep regime and the value of $\tau_m$ is determined by the initial condition, while the full fluidization process, characterized by $\tau_f$, corresponds to a process of diluting the memory of the initial state through multiple plastic activations. 

\subsection*{Aging dependence of the fluidization time scaling}

\begin{figure}[th]
\begin{center}
\includegraphics[width=\columnwidth, clip]{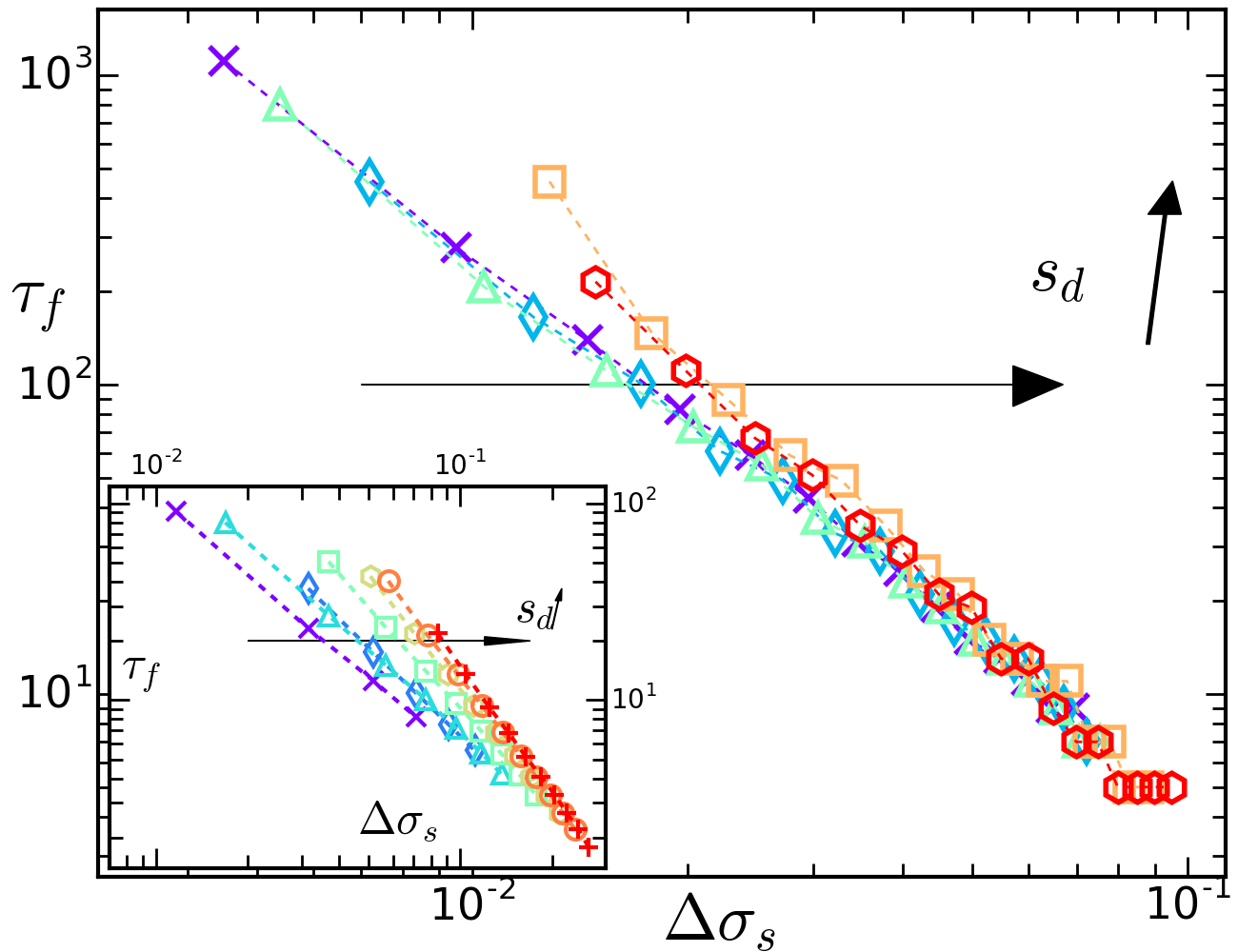}
\end{center}
\caption{
Main figure: $\tau_f$ v.s. $\Delta\sigma_s$ for the elasto-plastic model.
Inset: $\tau_f$ v.s. $\Delta\sigma_s$ for the mean-field model, $\alpha=0.3$. The values of $s_d$ are the same as in Fig.\ref{fig:tf}
}
\label{fig:tau_f_sigma_static}
\end{figure}

The previous analysis suggests that the critical yield stress above which the fluidization takes place, is not determined by the flow curve, but rather by the initial  state and its level of  relaxation. Hence, the divergence of $\tau_f$ as a power-law cannot be expected if the dynamic yield stress $\sigma_Y$ is taken as a reference.
We then estimate a static yield stress from the divergence of the fluidization time $\tau_f$, by identifying for each $s_d$ the value $\sigma^s_Y(s_d)$ for which a power-law $\tau_f \sim \Delta\sigma_s^{-\beta_s}\equiv [\sigma^\mathrm{ext}-\sigma^s_Y(s_d)]^{-\beta_s}$ holds.
Finding the best power-law fitting we estimate both $\beta_s(s_d)$ and $\sigma^s_Y(s_d)$.
The result of this analysis is shown in Fig.~\ref{fig:tau_f_sigma_static}. 
We observe power-laws that span over a decade each, with effective exponents depending on $s_d$. 
The exponent values display a  clearer trend in the mean-field case, where $\beta_s$ seems to increase systematically with $s_d$.  We note however that, for the elasto-plastic model, finite size effects may delay a bit the fluidization $\tau_f$ when the imposed stress is small, so that the static yield stress may be overestimated \footnote{The maximum system size for the simulations has been chosen such that we enter the stationary state after a reasonable running time. Performing a finite size analysis to more accurately estimate the value of static yield stress is difficult within our numerical protocol.}. 

\begin{figure}[b!]
\begin{center}
\includegraphics[width=\columnwidth, clip]{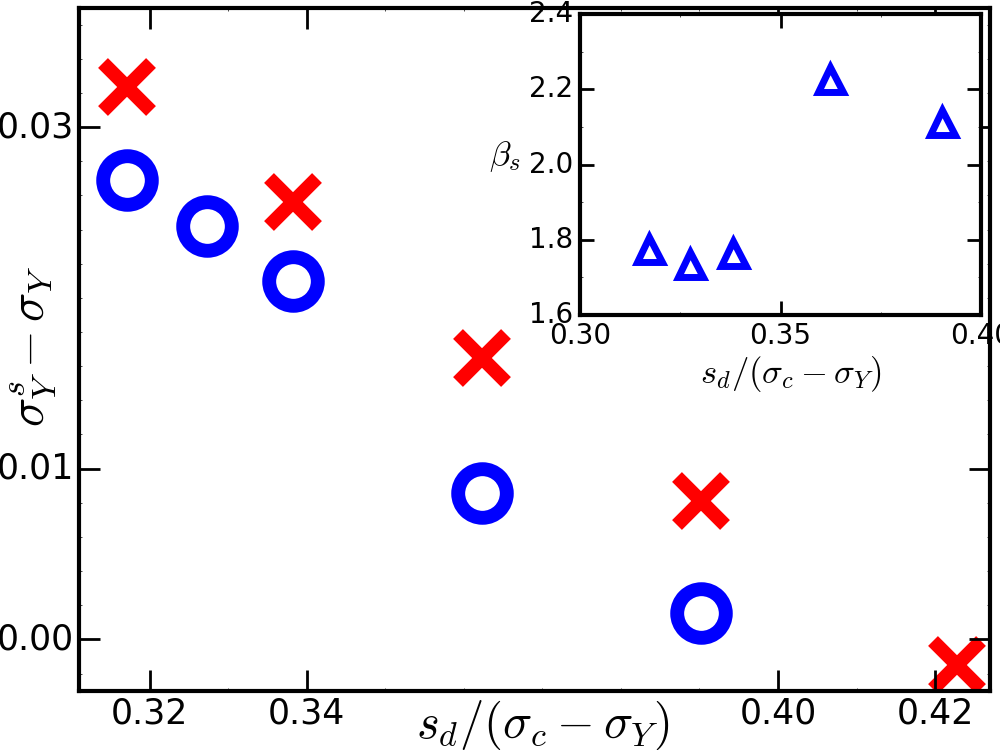}
\end{center}
\caption{
\textit{Elasto-plastic model}.
Main figures:  Circle: difference between static and dynamic yield stress $\sigma^s_Y-\sigma_Y$ against the relative relaxation coefficient $s_d/(\sigma_c-\sigma_Y)$. Cross: $\sigma^s_Y-\sigma_Y$ measured as the stress overshoot at zero shear rate limit. Error bars within the size of markers.
Insets: $\beta_s$ v.s. relative relaxation coefficient.}
\label{fig:sigma_static_beta_static_ep}
\end{figure}

\begin{figure}[t!]
\begin{center}
\includegraphics[width=\columnwidth, clip]{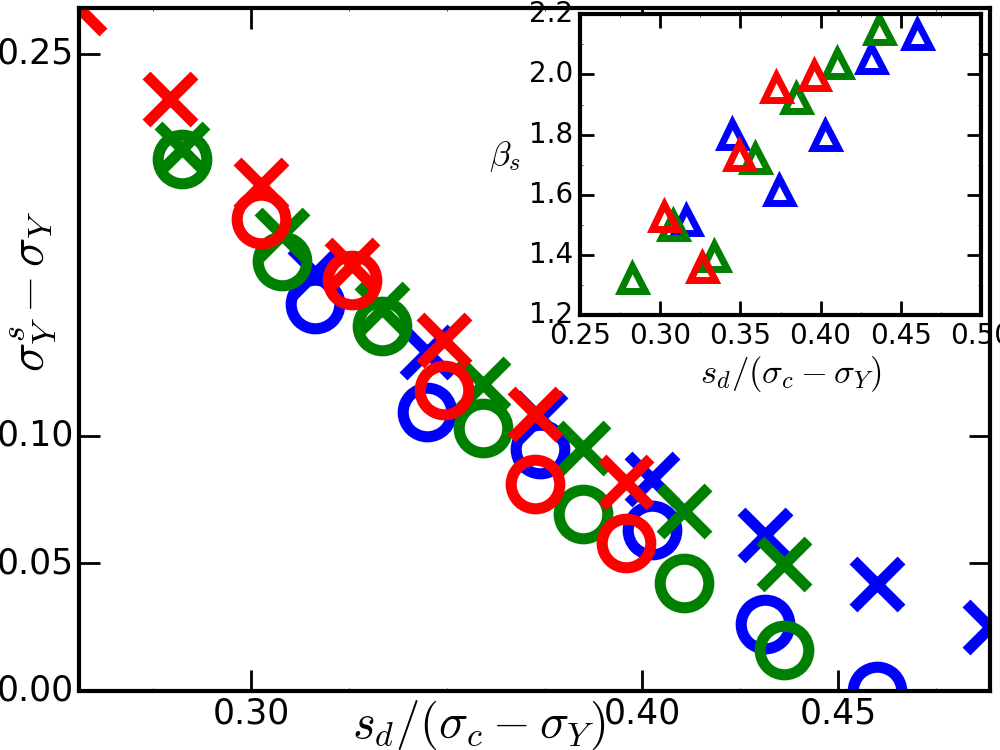}
\end{center}
\caption{
\textit{Mean-field model}.
Main figures: Circle: The difference between static and dynamic yield stress $\sigma^s_Y-\sigma_Y$ against the relative relaxation coefficient $s_d/(\sigma_c-\sigma_Y)$. Cross: $\sigma^s_Y-\sigma_Y$ measured as the stress overshoot at zero shear rate limit.
Insets: $\beta_s$ v.s. relative relaxation coefficient.
Blue: $\alpha=0.2$, Green: $\alpha=0.3$, Red:$\alpha=0.4$.
}
\label{fig:sigma_static_beta_static_mf}
\end{figure}

In order to make more explicit the dependency of $\sigma^s_Y$ and $\beta_s$  on $s_d$, we plot $\sigma^s_Y-\sigma_Y$ and $\beta_s$ against $s_d/(\sigma_c-\sigma_y)$ in Figs.~\ref{fig:sigma_static_beta_static_ep} and~\ref{fig:sigma_static_beta_static_mf} (circles). 
This choice of axes responds to the fact that in the mean-field model $\sigma_Y$ depends inversely on the mechanical coupling strength $\alpha$; so, the effect of the initial stress distribution are only comparable for different values of $\alpha$ when they are measured relative to the distance between the instability threshold $\sigma_c$ and the dynamical yield stress $\sigma_Y(\alpha)$.
In a way, the ratio $s_d/(\sigma_c-\sigma_Y)$ compares the relaxation level (or stress spread) prior to a creep test to the spread for a system flowing very slowly, since $\sigma_c - \sigma_Y$ characterizes the spread of the stress distribution in the zero strain rate limit.
When $s_d$ and $\sigma_c-\sigma_Y$ become comparable, we would expect $\sigma^s_Y-\sigma_Y$ to approach zero.
In other words, if the initial aging is not enough (large values of $s_d$), the overshoot in the stress-strain curve (that distinguishes $\sigma^s_Y$ and $\sigma_Y$) should cease to be observed. This is consistent with recent observations in particle based simulations \cite{Ozawa201806156} and confirmed by our data.
Both models show $\sigma^s_Y-\sigma_Y$ decreasing to zero as $s_d/(\sigma_c-\sigma_Y)$ approaches $\mathcal{O}(1)$. The insets of Figs.~\ref{fig:sigma_static_beta_static_ep} and~\ref{fig:sigma_static_beta_static_mf} present the creep exponent $\beta_s$ against the relative relaxation. 
We see, for both models, an increase in $\beta_s$ with increasing level of relative relaxation. Apart from some numerical fluctuations, the collapse of $\beta_s$ obtained  for different values of $\alpha$, shown in the inset of Fig.~\ref{fig:sigma_static_beta_static_mf}, suggests a master relation between $\beta_s$ and $s_d/(\sigma_c-\sigma_Y)$. 
The value of $\beta_s$ found at large $s_d$ is comparable with experimental measurements \cite{divoux2011stress}. We also plot $\beta$ against the relative relaxation level $s_d/(\sigma_c - \sigma_Y)$ in Fig.\ref{fig:beta}(b). Curves for different values of $\alpha$ collapse together, while the curve of the elasto-plastic model is a bit shifted aside.

\subsection*{The static yield stress}

Previously we have mentioned the relation between the static yield stress $\sigma^s_Y$ and the zero strain rate limit of the stress overshoot, we now address this point.
We switch to the strain rate controlled protocol and perform shear start-up simulations at different values of strain rate from the same initial conditions used for the creep tests.
We record the largest stresses reached in the stress-strain curves produced under different values of strain rates and initial relaxation levels ($s_d$). We find that for a given $s_d$ the stress overshoot decreases with applied strain rate, converging to a finite value when the strain rate goes to zero (see the supplemental material of Ref.\cite{LiuPRL18}).  
The corresponding limit values for each $s_d$ are plotted with crosses in Figs.~\ref{fig:sigma_static_beta_static_ep} and~\ref{fig:sigma_static_beta_static_mf}.
We observe that, although the static yield stress and the stress overshoot do not perfectly match, they stay representative  each other along the trend as function of $s_d$.
Comparing the spatial and the mean-field models, these quantities have a better agreement for small values of $s_d$, which justifies the idea that the underlying physics of the static yield stress observed in creep experiments and that of the quasi-static shear stress overshoot are closely related.
We also notice that at large values of $s_d/(\sigma_c-\sigma_Y)$, a small but systematic deviation exists.
In such situations, the initial conditions correspond to poorly relaxed systems, with a very short fluidization time, for which deviations from the scenario outlined above may be expected.

\subsection*{Correlations and cooperativity in the creep dynamics}

The effective dynamics described by Eqs.\ref{eq:hl_pq} and \ref{eq:hl_pa} and the results shown in Fig.~\ref{fig:shearrate_q} explain, from a mean-field perspective, in terms of populations above and below characteristic values, the underlying physical process involved in creep and fluidization.
In the spatially resolved model we are able to further see the spatial distribution of these  populations.
Therefore, we discuss now the underlying physics of creep dynamics from the point of view of the cooperativity of local plastic events in the elasto-plastic model.
Let us recall that on each site, the state variable $n_{ij}(t)$ alternates between zero and one respectively for elastic and plastic states.
In the creep simulation, one sub-volume or block contributes to the macroscopic strain rate only when in its plastic state so that the time evolution of the state variable field can be used to infer the physical process underlying creep.

In order to implement this, we choose to accumulate the state variable during a given time window
\begin{eqnarray}
f_{ij}(t,\Delta t) = \int_{t-\Delta t/2}^{t+\Delta t/2} n_{ij}(t')dt'
\end{eqnarray}
The field $f_{ij}$ integrates the plastic activation information during that time window. Here $\Delta t$ is chosen to be of the same order of magnitude as $t$ for the plastic activations at a given time scale $t$ to be well represented by $f_{ij}(t)$. (Actually we have $\Delta t\approx 2t/3$.)  We then compute the following correlation map
\begin{equation}
C_{\Delta i, \Delta j}(t)= \frac{\langle f_{ij} f_{i+\Delta i,j+\Delta j}\rangle-\langle f\rangle ^2}{\langle f^2\rangle - \langle f\rangle ^2}
\end{equation}
where $\langle\cdot\rangle$ represents ensemble averages.
To characterize a ``cooperativity level'' of the plastic events taking place within a corresponding time window, we define the correlation intensity $I_C$ as the integral of the absolute value of the correlation map:
\begin{equation}
I_c(t) = \int \big|C_{\Delta i, \Delta j}(t)\big|dv .
\end{equation}
This quantity is indicative of how strongly plastic events are correlated during a specific stage of the creep test. 

\begin{figure}[t!]
\begin{center}
\includegraphics[width=\columnwidth, clip]{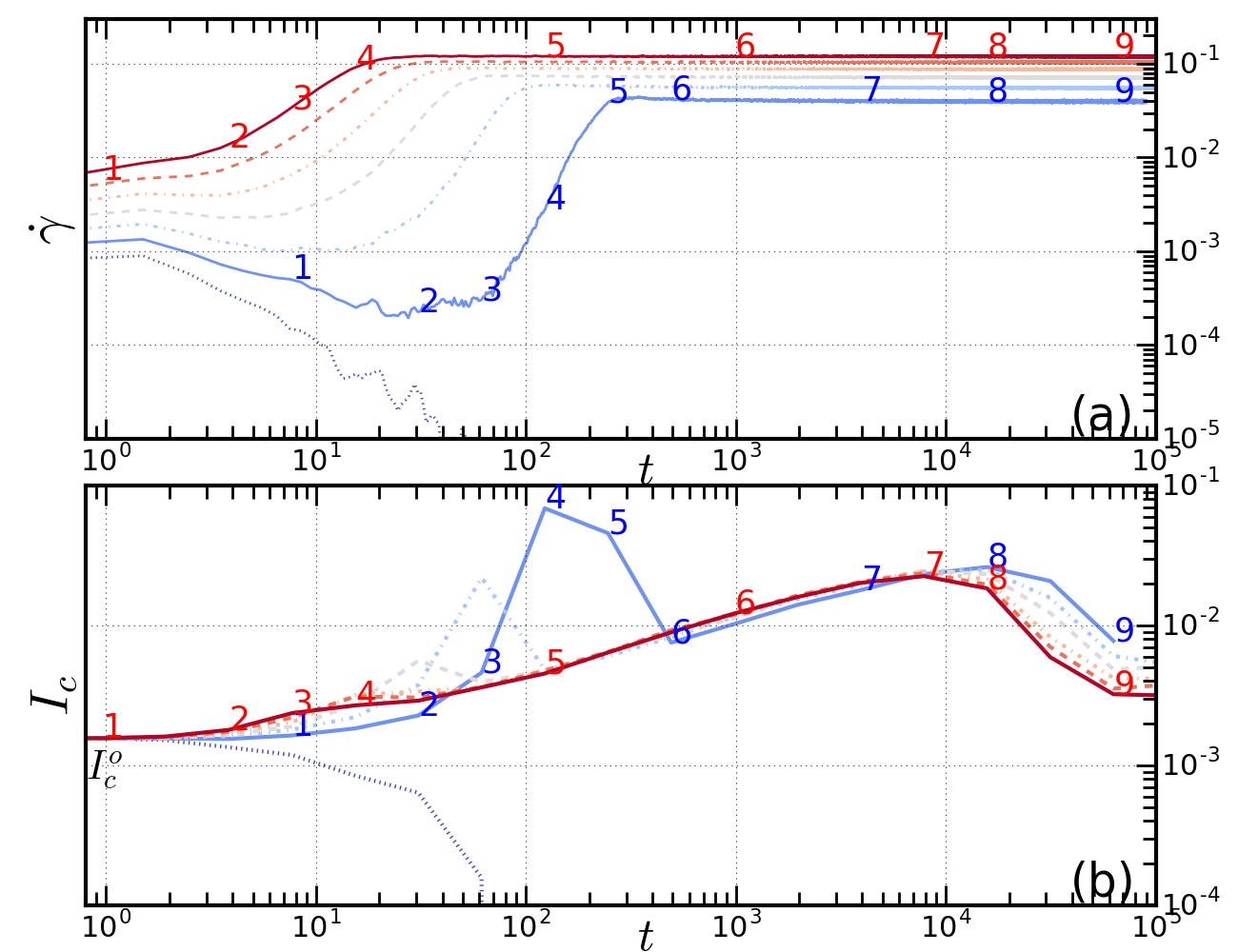}
\end{center}
\caption{Elasto-plastic model: The strain rate $\dot{\gamma}(t)$ (a) and the corresponding correlation intensity $I_c(t)$ (b) for different applied stresses. From bottom to top $\Delta \sigma = 0.025,\;0.035,\; \ldots,\; 0.85$}
\label{fig:correlation_intensity}
\end{figure}
The correlation intensity as function of time and the corresponding $\dot{\gamma}(t)$ for different values of the imposed step stress are presented in Fig.~\ref{fig:correlation_intensity}. 
After a linear increase of both $I_c(t)$ and $\dot{\gamma}(t)$ for $t\lesssim t_{mic}\simeq 1$ (not-shown), leading all curves to a common  $I_c^o$ level, we observe distinctive features according to the value of the stress.
For the smallest applied stress, the correlation intensity decreases with time and drops to zero, similar to the strain rate.
For stresses that lead to steady state flow in the long term, the correlation intensity continues to increase beyond  $I_c^o$. Note that while the behaviour of the strain rate at early stages is not a clear indicator of fluidisation, an increasing $I_c(t)$ appears to be clearly correlated with it.

For those curves leading to a steady flowing state in Fig.~\ref{fig:correlation_intensity}(a), the corresponding correlation intensities $I_c(t)$ (Fig.~\ref{fig:correlation_intensity}(b)) exhibit two characteristic time scales: 
(1) a bump of $I_c$ before entering the steady state flow, for example between points \textbf{3} and \textbf{6} on the blue curve and between points \textbf{2} and \textbf{4} on the red curve.
This bump lies roughly in the same range as the inflection point in $\dot{\gamma}$, i.e. corresponds to the fluidization time $\tau_f$.
This relates the fluidization time scale $\tau_f$ to a maximum cooperativity  of plastic events.
The amplitude of  the  maximum in $I_c$ decreases with increasing the applied stress beyond $\sigma^s_Y$.
When the applied stress is big enough, less cooperation is needed in the system in order to overcome the static yield stress.
The large amplitude in $I_c$ observed between points \textbf{3} and \textbf{6} for the blue curve corresponds to a stress that is very close to the static threshold $\sigma^s_Y$. 
There, significant spatial correlations are needed in order to fluidize the system.
(2) At longer times, after entering the steady state where no other time scales can be recognized from the $\dot{\gamma}(t)$ curves, further information comes from the correlation intensity.
$I_c(t)$ displays a local maximum at long time (point \textbf{7} on the red curve) after which it decreases again to a low value in the steady state, point \textbf{9}.
Although the strain rate seems to reach a steady state earlier, the true steady flowing state is achieved only after the correlation level (cooperativity) is relaxed to a steady value.
In fact, the strain rate is still evolving up to that moment, but on a very small scale.
One recognizes also in Fig.~\ref{fig:correlation_intensity} that the steady state cooperativity level of plastic events is inversely related to the steady state strain rate.
Intuitively, a fast strain rate tends to activate plastic events rather randomly, while a slow strain rate leaves enough room (time) for correlations in plastic activity  to develop \cite{liu_thesis}. 

\subsection*{Activity maps and spatial correlation maps}

\begin{figure}[t!]
\begin{center}
\includegraphics[width=0.9\columnwidth]{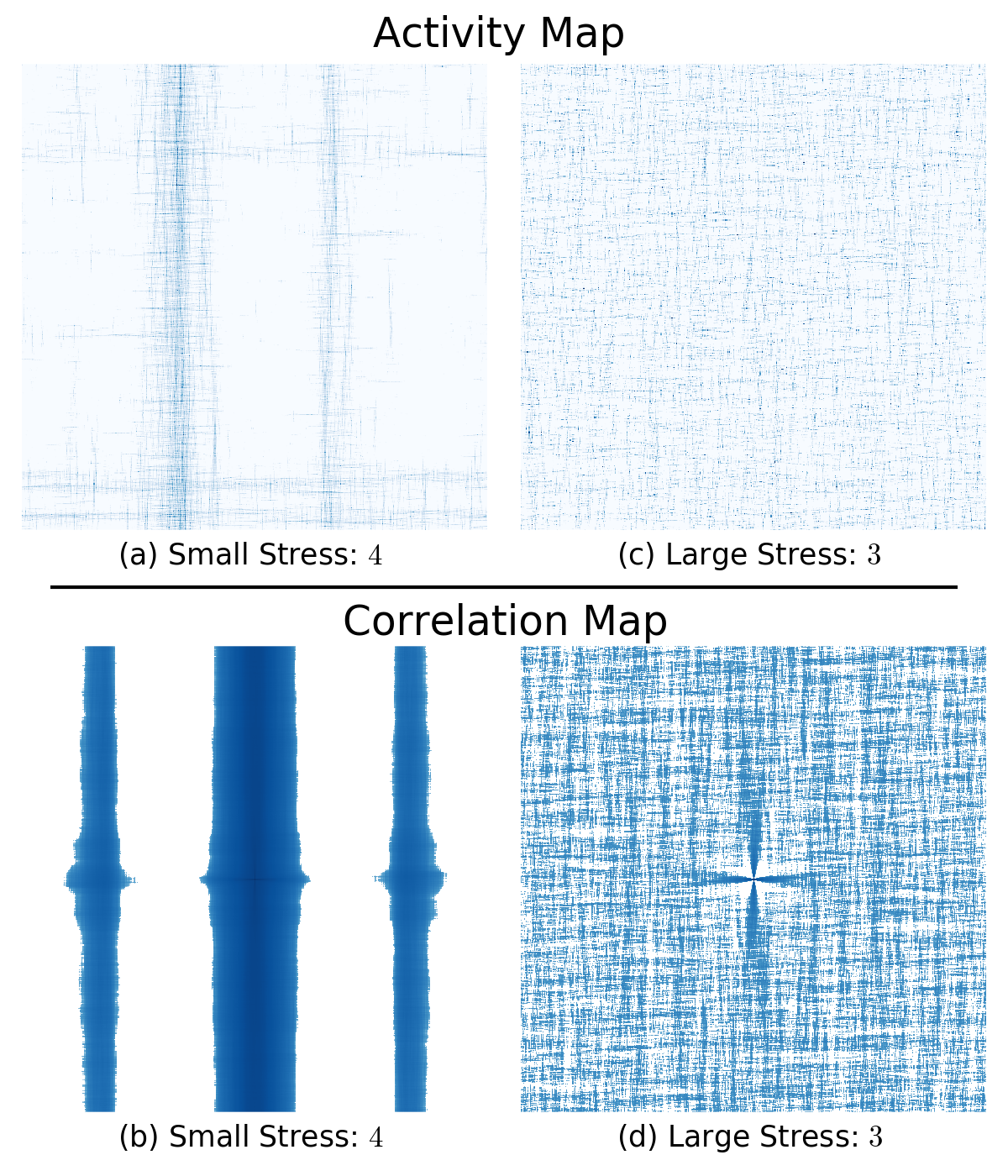}
\end{center}
\caption{
Accumulated plasticity maps and correlation maps within a time window for the first peak in the correlation intensity. Correlations are plotted in log scale for a better visibility. (a)(b) correspond to the point $4$ of the blue solid curve in Fig.\ref{fig:correlation_intensity}. (c)(d) correspond to the point $3$ of the red solid curve in Fig.\ref{fig:correlation_intensity}.
}
\label{fig:active_corr1}
\end{figure}

The previous discussion of the development and correlations of cooperativity can be explicitly visualized in our spatial model by  snapshots of accumulated plasticity and spatial correlations within a time window, at different stages of the dynamics and for different applied stresses. Figs.~\ref{fig:active_corr1} shows these maps at the instants marked by the number labels in Fig.~\ref{fig:correlation_intensity}. 

At the beginning, not far from $I_c(t)=I_c^o$, even when there is significantly more plastic activity for the larger applied stresses than for the smaller one, their correlation maps are almost structureless. 
A clearer spatial structure of correlation only appears at the onset of fluidization, close to the peak of  $I_c(t)$ (Fig.~\ref{fig:active_corr1}(b,d)).
At this point, the accumulated activity looks homogeneous in space for the large applied stress (Fig.~\ref{fig:active_corr1}(c)), while the cooperativity of plastic events is confirmed by the quadrupolar form in the correlation map displayed in Fig.~\ref{fig:active_corr1}(d).  On the other hand, for the small applied stress we already observe the plastic activity organized in several vertical and horizontal bands (Fig.~\ref{fig:active_corr1}(a)) and a very pronounced vertical correlation pattern in Fig.~\ref{fig:active_corr1}(b), corresponding to the fluidization phase (\textbf{4} to \textbf{5} in Fig.~\ref{fig:correlation_intensity}).
This suggests that a strong correlation developing along the y-direction could be responsible for the burst of plastic events leading to fluidization (and the speed-up of $\dot{\gamma}$).
Of course, there is no a priori preferred direction for cooperativity and the vertical correlation band observed in this example should be equally frequent as an horizontal one. Note that the elasto-plastic model may over-emphasize the stripes formation for the cases of small applied stresses due to the Fourier space implementation of the interaction kernel $G$ and the fact that the model assumes homogeneous elasticity mediating the interactions among plastic events. Nonetheless, transient shear banding during creep is indeed observed in more realistic MD models\cite{chaudhuri2013onset}.

After the fluidization regime, the behavior of plastic activity for the smaller applied stress is similar to that for the larger applied stress (not shown).
For the large applied stress, the quadrupolar correlation gradually develops until one arrives at the last maximum in $I_c$  and drops back to a very weak spatial pattern in the steady state.
Correspondingly, plastic activity is more organized into thin slip lines at the last peak of the correlation intensity than in the steady state.
The phenomenology is a bit more complex for the small applied stress, close to $\sigma^s_Y$.
The burst of cooperative plastic events that organize into the shear bands in the fluidization regime not only speeds up the strain rate but may also induce  sites outside the bands to become unstable: since plastic activations inside a band tend to decrease the stress, this should be balanced by an increase of local stresses outside the bands in order for the overall averaged stress to keep constant.
As a result, more plastic events take place randomly outside the shear bands; this acts against overall cooperativity: the correlation intensity drops significantly after the large bump (\textbf{3} to \textbf{6} in Fig.~\ref{fig:correlation_intensity}). The remaining process afterwards is very much like the one in the case of the large applied stress. Activity maps show thin random slip lines for both stresses and the correlations are both very weakly quadrupolar.
The only difference is that the last maximum in $I_c(t)$ appears later and the final stationary correlation pattern is a bit more pronounced for the smaller stress. 

\section*{Conclusions}

We used both spatially-resolved and mean-field mesoscopic models to study the creep behavior of athermal amorphous materials with different initial relaxation degrees.
Despite the simplicity of the models, they are sufficient to reproduce the \textit{S}-shaped strain rate response observed in experiments. 
Further, the two models are consistent in 
both qualitative and quantitative manners, so that the mean-field model can be considered as a simplified way to understand the more realistic spatially resolved model.
We measured the power law slowing down $\dot{\gamma}\sim t^{-\mu}$ in the creep regime.
We found that the exponent $\mu$ produced by both models lies in the same numerical range as experimental results but is not universal with respect to the applied stress and the level of initial relaxation.
We distinguished, within the framework of the models, the different underlying physical processes for the two time scales $\tau_m$ and $\tau_f$ that characterize the creep behavior.
$\tau _m$ is determined by the initial stress distribution $P_0(\sigma)$ around the marginal stability threshold $\sigma_c$, while $\tau_f$ is closely related  to subsequent plastic activations and spatial cooperativity of plastic events.
We interpreted our results on the relation between the fluidization time $\tau_f$ and the applied stress $\sigma^{EXT}$ by defining a static yield stress $\sigma^s_Y$, which increases with initial relaxation.
A convincing power law $\tau _f\sim(\sigma^{EXT}-\sigma^s_Y)^{-\beta_s}$ is observed, with $\beta_s$ increasing when shortening the initial aging and taking values comparable to those reported in experimental studies.
Finally, we defined an intensity of spatial cooperativity that can serve as a precursor to distinguish systems that fluidize from those stuck at the early stage of the creep phase. 
The onset of the fluidization regime  is associated, especially for small stresses, with a strong spatial cooperativity.
Moreover, we noticed that spatial correlations in cooperativity seem to be qualitatively different between systems that undergo  a creep regime prior to the fluidization and those that fluidize directly.

\section*{Acknowledgements}
J.-L.B. and C.L. acknowledge financial support from ERC grant ADG20110209.
K.M. acknowledges financial support of the French Agence Nationale de la Recherche (ANR), under grant ANR-14-CE32-0005 (FAPRES) and of the Centre franco-indien pour la Promotion de la Recherche avanc\'ee (CEFIPRA) Grant No. 5604-1 (AMORPHOUS-MULTISCALE).
Authors acknowledge support from the collaboration project ECOS Sud-MINCyT A16E01. 

\bibliography{biblio_creep2} 

\end{document}